\documentclass[12pt, A4paper]{article}
\usepackage{caption}
\usepackage{subcaption}
\usepackage{amsmath}
\usepackage{verbatim}
\usepackage{setspace}
\usepackage{rotating}
\usepackage{pdflscape}
\usepackage{morefloats}
\usepackage{hyperref}
\usepackage{graphicx, xcolor}
\usepackage{color, colortbl}
\usepackage{url}
\usepackage{amssymb}
\usepackage[sectionbib]{natbib}
\usepackage{sectsty}
\usepackage{multirow}
\usepackage{longtable}
\usepackage{rotating}
\usepackage[labelfont=bf,font=footnotesize]{caption}
\usepackage[left=2.56cm,right=2.56cm,top=2.56cm]{geometry}
\usepackage{mathrsfs}
\usepackage{dsfont}
\usepackage{enumerate,letltxmacro}
\usepackage[normalem]{ulem}
\LetLtxMacro\itemold\item

\usepackage[english]{babel}
\usepackage{blindtext}
\definecolor{MyGreen}{cmyk}{1.0,0.0,1.0,0.2}

\definecolor{Gray}{gray}{0.9}
\usepackage[font=singlespacing]{caption}

\begin{document}
\clearpage
\setcounter{page}{1}
\title{Measuring and Assessing Latent Variation in Alliance Design and Objectives\footnote{The author thanks Janet Box-Steffensmeier, Bear Braumoeller, Olga Chyzh, Skyler Cranmer, Marina Duque, Christopher Gelpi, Jared Edgerton, Andrew Goodhart, Daniel Kent, Shahyar Minhas, Mark Nieman, Paul Poast, faculty and students of the University of Wisconsin, Purdue University, Wayne State University, Northeastern University, and Dartmouth University, and participants of the Research in International Politics Workshop for their thoughtful comments.}}
\author{Benjamin W. Campbell\footnote{BWC: Doctoral Candidate in Political Science, The Ohio State University, e: \href{mailto:campbell.1721@osu.edu}{campbell.1721@osu.edu}}}
\date{\today}
\maketitle
\vspace{1cm}


\begin{abstract} 
\noindent  
The alliance literature is bifurcated between an empirically-driven approach utilizing rigorous data, and a theoretically-motivated approach offering a rich conceptualization of alliances.  Within the strength of one, lays the weakness of the other.  While the former invokes a non-comprehensive view of alliances that emphasizes capability aggregation, the latter does not provide a systematic or rigorous approach to uncover empirical insights.  I unify these perspectives, enumerating the roles a state can adopt within the alliance network and considering the relationship between that role and how they design their local alliance network to accomplish role-based objectives.  To uncover this variation, I employ a novel methodological tool, the ego-TERGM.  Results indicate that states form alliances to accomplish a variety of foreign policy objectives beyond capability aggregation, including the consolidation of non-security ties and the pursuit of domestic reforms in addition to security-based motivations.  
\end{abstract}

\clearpage
\doublespacing
\section{Introduction}
Few international institutions are as ubiquitous as military alliances.  
With far reaching consequences, interstate military alliances influence almost every aspect of international politics. 
Even the foundational theory of international relations, balance of power theory, is a theory about when states should aggregate their capabilities through alliances.  
However, for a concept so integral to so many international processes and research areas in international relations, little attempt has been made to empirically examine variation in why states form alliances.  
Since the foundation of international relations, alliances have been considered a means of promoting state security through deterring attack, balancing competitors, or preparing for an inevitable conflict \citep{carr1939twenty, hans1948politics, waltz1979theory, mearsheimer2001tragedy, johnson2017external}.   Empirical approaches have largely mirrored this understanding \citep{lai2000democracy, leeds2002alliance, leeds2003alliance, gibler2006alliances}, while a variety of theoretically-motivated studies have pushed back, recognizing the variety of motives states may have for forming alliances \citep{morrow1991alliances, schroeder1994historical, schroeder1996transformation, kim2016supply, henke2017politics}.  Recent empirical work has highlighted that alliances may vary in their design \citep{leeds2002alliance}, but the initial conditions motivating their formation are thought to be relatively homogenous.  While theoretically-motivated scholars have been increasingly interested in empirically considering variation \citep{kim2016supply, henke2017politics}, it has been largely been limited to particular case studies.  

The previous discussion highlights two distinct approaches dominating the study of military alliances. The peace science approach is empirically-driven and uses systematic and rigorous data to uncover meaningful empirical insights about the causes and consequences of alliances.  The security studies approach is more theoretically motivated, using context and diplomatic history to provide a rich and nuanced conceptualization of alliances.  Unfortunately, these approaches rarely engage with one another, and within the strength of one lays the weakness of the other.  While the peace science approach yields tractable empirical insights, it does so by invoking a flat and non-comprehensive view of alliances.  This view emphasizes the Capability Aggregation Model (CAM) which holds that states, motivated by an external security threat, form alliances to aggregate capabilities and consolidate their security.  Many security studies scholars have long voiced their opposition, arguing that alliances are complex institutions, and states form alliances for a variety of motives.  While the security studies approach provides a means of understanding this nuance, generating inferences is difficult as empirically assessing and generalizing this variation is difficult.  In this manuscript, I unify these perspectives, answering several questions:  \textit{Are alliances only formed to respond to external threats?  Do contextual factors influence states' motives for forming alliances? How do we synthesize the parsimonious Capability Aggregation Model with contextual factors? How do the objectives motivating alliance formation differ from one another, and how would we measure these differences?  Is this desire to differentiate alliances analytically useful?}  


Synthesizing the peace science and security studies approach, I introduce a new role-based framework for considering variation in alliance design and objective.  
This role-based framework holds that states adopt roles within the alliance network that vary across systemic context, and design their local alliance network to accomplish role-motivated objectives.  
These roles are thought to vary across two dimensions -- whether they are order pursuing/preserving or narrowly utilitarian, and whether the aggregation of capabilities is the chief means to accomplish their objectives or if institutional means are essential.
States may adopt the role of Balancer (Ordering, Capability Aggregating), Aggregator (Narrowly Utilitarian, Capability Aggregating), Reformer (Ordering, Institutional), or Consolidator (Narrowly Utilitarian, Institutional), each possessing its own logic and objectives motivating local alliance network design.  
These roles are uncovered using the ego-Temporal Exponential Random Graph Model (ego-TERGM), a novel statistical innovation that allows analysts to sort ``ego-networks" (e.g. a state and their allies) into a finite number of clusters according to similarity.  
This model allows for both the generation of data on roles and an assessment of the generative process for these roles.  

By synthesizing the richness of the security studies approach with the rigor of the peace science approach, I find that the Capability Aggregation Model of alliances is inadequate and that scholars should empirically consider the heterogeneous and non-security motives driving states to form military alliances.  
While security-based roles are detected, states are frequently found to form alliances to consolidate economic or political relationships and to promote domestic reforms.   
The roles uncovered are also found to vary according to historical context.
This novel approach has far reaching implications for how IR scholars consider alliance formation and evolution, and the consequences of different alliance types. 

\section{State Objectives Influencing Alliance Design}
The literatures on alliance formation and maintenance span two schools with two distinct approaches:  peace science and security studies.  
The conventional wisdom in peace science, and international relations broadly, is embodied by the parsimonious and generalizable Capability Aggregation Model (CAM).  The CAM holds that states form and maintain alliances to balance peer competitors, deter potential attack, improve bargaining power, or prepare for what is perceived to be an inevitable conflict \citep{johnson2017external}.  This perspective has dominated the empirical literature on alliances as its parsimonious logic makes modeling alliance behavior simple and tractable using existing data sources.
While such motivations inform many alliances, there remain cases that this logic cannot explain, including the enlargement of the North Atlantic Treaty Organization (NATO) \citep{waltz2000structural}, the post-Cold War evolution of the Organization of American States (OAS) \citep{pevehouse2005democracy, therien2012changing}, or the formation of the Economic Community of West African States (ECOWAS) \citep{ukeje2005economic}.  
Security studies and diplomatic history, on the other hand, offers nuanced explanations for why particular alliances form and evolve without emphasizing generalizable or empirically-oriented logics capable of detecting such patterns.  

In other words, within the \textit{strength} of one approach lies the \textit{weakness} of the other.  
While international relations and peace science possesses a generalizable and parsimonious model of alliances that travels to many cases and makes empirical modeling easy, it fails in explaining a great many important alliances.  
While security studies and diplomatic history can offer nuance and explain the contexts driving particular alliances, it lacks a general framework capable of detecting consistent logics.  In this manuscript I argue that a synthesis of these approaches is necessary to further theoretical and empirical innovation.  

\subsection{Alliances as Responses to External Threats}
The peace science approach, the dominant perspective within international relations, holds that states, motivated by some common external threat aggregate their capabilities to increase their security, bargaining power, and/or war-fighting capabilities beyond what they would be in the absence of a formal obligation alone \citep{morrow2000alliances, johnson2017external}.  This perspective is attractive for empirical modeling as it allows the analyst to assume there is a homogenous data generating process.\footnote{This perspective was certainly not created by peace scientists, nor is it only adopted by peace scientists or the only perspective within peace science.  Regardless, I label this perspective the peace science approach as it is the modal perspective in peace science, which is one of the most active literatures and most empirically-oriented literatures on alliances.}  
This view was propagated by early Realists and has dominated international relations since its founding.

In his call for a scientific study of international politics, \citet{carr1939twenty} critiqued Interwar foreign policymaking which prioritized ideals to a pursuit of the national interest.  
This view, formalized by \citet{hans1948politics}, identifies national power as the chief motive of states.  
Implicit to this logic is that states seek power to secure their sovereignty from external threats, and that at times, alliances may be necessary to ensure security.   
While many have debated the precise conditions necessitating such power politics and alliance decisions \citep{waltz1979theory, walt1990origins, mearsheimer2001tragedy}, the consequence is the same: states form alliances to secure their survival against external threats.  
Once these threats are removed, their corresponding alliances should dissolve \citep[904]{morrow1991alliances}.

With the quantitative shift in international relations and the rise of peace science, the empirical work on alliances has largely adopted these parsimonious assumptions.  Invoking this assumption makes modeling challenges much more tractable as analysts can assume there is a homogenous data generating process that is easily measurable.  
Early quantitative work on alliances emerged from the Correlates of War (COW) Project which coded alliances according to whether the alliance was a defense pact, a neutrality and non-aggression pact, or an entente \citep{singer1966formal}.  
These data have been used to examine many questions, including the factors influencing alliance formation \citep{lai2000democracy} and whether alliances deter or provoke conflict \citep{singer1966formal, smith1995alliance, gibler2006alliances}.  Building upon this dataset, \citet{leeds2002alliance} introduced the Alliance Treaty Obligations and Provisions (ATOP) dataset which included additional data on aspects of alliance design.  These data represent the state of the art and have allowed for a more rigorous examination of the factors underlying alliance formation \citep{leeds2002alliance, cranmer2012toward, cranmer2012complex}, and its effects on conflict \citep{leeds2003alliances}, trade \citep{long2006trading, fordham2010trade} and more \citep{benson2016assessing}.\footnote{There are many other well known and widely cited articles I would like to mention here, but journals have word limits.}  

It is not the COW or ATOP coding decisions that are problematic, but rather an analyst's comfortability in assuming a homogenous data generating process, treating alliances as varying only in commitment and not objective.  There are several challenges to this approach.
First, Diplomatic Historians have long noted that the empirical record challenges the ``no functional differentiation" assumption, arguing that states perform highly specialized functions \citep{schroeder1994historical}.  
Consider the OAS, an alliance formed to counter Soviet expansion into the Western Hemisphere, only to update its primary mission following the Cold War -- to accomplish a series of non-security goals including political and economic liberalization \citep{pevehouse2005democracy, therien2012changing}.  
Second, world politics reflects a highly dynamic system where a variety of time varying material and immaterial forces influence state behavior.  
Failing to model alliances as a function of this dynamic system may produce flawed inferences, as noted by 
\citet{jenke2016theme}.  
As such, treating alliances and their formational logic as time invariant may produce flawed inferences. 

Unit and temporal heterogeneity in alliance decision-making has certainly been explored by IR scholars, but never in a systematic way that posits a framework capable of competing with the CAM, leaving much of the work on alliance heterogeneity to diplomatic history and security studies.
\citet{holsti1970national} enumerated the foreign policy roles states adopted during the Cold War, which almost certainly influenced their motives for forming alliances.   
An early piece by Russett used factor analysis to cluster alliances on a variety of background and output variables to produce an alliance typology based upon an alliance's distribution of capabilities \citep{russett1971empirical}.  
In a recent study \citet{benson2016assessing} consider variation in the commitments enshrined in treaties, but does not consider the motivations for distinct commitments.  
The effect of alliances on conflict are also thought to differ before and after the advent of nuclear weapons \citep{kenwick2015alliances}.  
Existing understandings of alliance heterogeneity have failed to articulate a theoretical or empirical framework capable of explaining why alliances are formed or evolve to accomplish objectives beyond those outlined by the CAM.  

\subsection{Alliances as Heterogenous Institutions}
While diplomatic historians have certainly championed balance of power theory \citep{taylor1954struggle}, the modal approach holds that alliances possess a great degree of specialization and are used to accomplish many objectives \citep{schroeder1994historical, kissinger1994diplomacy, schroeder1996transformation, bridge2014great}.  In his well known retort to \citet{waltz1979theory}, \citet{schroeder1994historical} argues that states have unique logics of political survival and have sought survival through performing specialized roles (125).  This call for nuance in treating alliances as institutions formed for a variety of reasons, while largely falling on deaf ears within mainstream international relations, has been recently heralded by security studies scholars.  

States have different logics of political survival and have different instruments for achieving them.  
This functional differentiation in the roles adopted and performed by states has significant consequences for how and why states form alliances.  
The objective of the United States following World War II was to contain the Soviet Union through establishing alliances in any area at risk of communist infiltration \citep{holsti1970national, kissinger1994diplomacy}.  
The \textit{Chief Balancer} role ascribed to and adopted by the United States was well understood and certainly transformed the US' incentive structure, driving it to design its local alliance network to be consistent with this role.  
Following the Cold War, the US is widely thought to adopt the new role as \textit{Hegemon}, a promoter of a global order based upon liberal institutions \citep{ikenberry2012liberal}.  
The US and its allies adjusted their existing alliance network accordingly, bringing former Soviet satellites into NATO and establishing a new mission for the OAS.  

This nuanced view is standard within diplomatic history and security studies, with scholars offering rich and sophisticated narratives for the historical context of each state, and how the position ascribed to them informs how they design alliances \citep{schroeder1994historical, schroeder1996transformation, ikenberry2012liberal, bridge2014great, kim2016supply, henke2017politics}.  
While many mainstream IR scholars or peace scientists might argue that the alliance system constructed by Metternich following the Concert of Vienna was an attempt to aggregate Russian, Prussian, and Austrian capabilities to prevent the return of Napoleon or to counter a very dominant United Kingdom \citep{jervis1992political}, many diplomatic historians are quick to recognize that the Holy Alliance was a pact among autocrats to protect sovereigns against rising nationalism and to promote a new order based upon cooperation \citep{schroeder1992did}.  Within security studies, theoretically-motivated work has explored variation in why states contract alliances.  In criticizing the CAM and its inability to explain alliances formed between great powers and weak powers, \citet{morrow1991alliances} introduced the Security-Autonomy Tradeoff Model which sought to explain the asymmetric gains of states forming alliances.  Relatedly, \citet{kim2016supply} has introduced a market theory of alliances, viewing alliances as a good exchanged between states wherein security may be exchanged for concessions in other areas.  Additionally, \citet{henke2017politics} explores how a variety of non-military factors influence a state's admittance to a multilateral military coalition.  While these are just examples from a much broader literature, a common thread has been the development of sophisticated and rich understandings and theories of alliances.  

Unfortunately, diplomatic historians and security studies scholars often lack parsimonious and generalizable frameworks for considering alliances or detecting patterns of alliance behavior.  They bring rich nuance to the study of alliances, often at the cost of producing systematic or rigorous empirical insights or predictions.  
As \citet{jervis1992political} notes, one large attraction of the IR approach to considering alliances is that it is relatively straightforward, parsimonious, and with a few simple assumptions it can explain a ``great deal of behavior that has taken place over tens of centuries under widely differing circumstances."  
This generalizability, a strength of the IR approach, is certainly a significant weakness of the diplomatic history approach to studying alliances.  

How might one synthesize these rich and context-based understandings of alliances with the generalizable models that interest IR scholars and peace scientists? In the following section I attempt to answer this question by introducing a new role-based framework for considering military alliances, combining the richness of diplomatic history and security studies with the empirically-motivated peace science tradition.  

\section{The Role Framework of Alliance Politics}\label{theory}
In this section a role-based framework for considering alliances is presented.  
The intention is not to explicitly theorize about why states adopt certain roles or to be exhaustive in enumerating the innumerable specialized roles a state may adopt.  Instead, the focus is to discuss what a ``role" is within the context of the alliance network, what these functionally differentiable roles might look like, what general patterns in roles might be detected, how they might be differentiated, and how they may vary as a function of time and system features.  


It is worth mentioning that the roles that actors adopt are undeniably dynamic and informed by systemic factors.  A broader discussion of this temporal heterogeneity and system-based expectation is included in the Supplementary Information (SI) Appendix.   I explicitly model this heterogeneity when later assessing this framework.  

\subsection{Roles and the Alliance Network}
Within international relations and political science broadly, the dominant understanding of roles stems from \citet{wendt1999social}.  For Wendt, a role is the cultural position an actor performs by occupying a place within a social structure and observing behavioral norms towards Others who perform a counter-role \citep[227]{wendt1999social}.    
Roles, as they are discussed in this manuscript, differ from Wendt's understanding.  Social roles, as discussed here, refer to cultural objects that are widely accepted and understood within a given community and used to accomplish certain objectives \citep{campbell2017detecting}.  This definition is similar to Wendt's in that roles are attributes of a broader system and are rooted in certain repeated and emergent practices.  While both understand roles as structural features, Wendt views them as mutually constitutive while I view them as purpose-based.



In other words, I believe that states have more agency in making these decisions.  In addition, role need not necessarily precede behavior, in some cases role may be adopted to conform to current behavior.  Within the context of the alliance network, a network defined by states related to one another through the presence of shared alliance treaties, states may adopt whatever roles their alters and the broader system will allow them to.  

This view of states stands in contrast to the strong assumption often invoked that states are not functionally differentiable, that they only differ with respect to relative capabilities \citep{waltz1979theory, schroeder1994historical}.  If states behave differently based upon observable roles, then the theoretical underpinnings for much of modern international relations theory must be reevaluated.  In the following section I give cause for such concern.

\subsection{Roles and Objectives for Alliance Formation}

	
	
	

As previously noted, and discussed in \citet{holsti1970national}, the CAM holds that states only form alliances to further objectives consistent with one of three roles: Balancer, Aggressor, or Defender.  
It is undeniable that across time and space, this capability-aggregation function of alliances is a constant.  
The stated role of the US and other NATO members as both Balancers and Defenders during the Cold War is clear \citep{holsti1970national}.  What is less clear is whether one can say that these roles persist when the supposed Aggressor, the USSR, dissolves.  
Without a clear threat or motive, the CAM cannot make sense of NATO's post-Cold War expansion \citep{morrow1991alliances, waltz2000structural}.  

To make sense of this puzzle, I introduce a role-based framework for considering alliances.  
This approach differs from the much narrower CAM by positing that states may have motives for adopting roles that are not simply motivated by countering external threats.  
The framework presented here holds that there are four ``types" of states in the alliance network that are characterized by the role adopted and their expected behavior.  
These roles are sorted according to two characteristics, forming the configuration presented in Table \ref{2by2}.

The first significant dimension that may sort alliances is whether the alliance is formed for some order-based objective or whether the alliance should be considered narrowly utilitarian \citep{lake2009regional, schweller2011after, ikenberry2012liberal, ikenberry2014power}.  When considering order, one must consider the role of ideology which certainly informs an actor's preferences with respect to ideal orders.  Within this context, the international orders pursued by states ``represent sets of governing arrangements, rules, and norms designed to secure the constituent units against external threat, at the cost of limiting their autonomy," \citep{braumoeller2019}.  The particular set of arrangement, rues, and norms are often an outcome reached through bargaining among states \citep{lake2009hierarchy}.  Alternatively, alliances may be pursued for narrowly utilitarian reasons, as is usually thought.  These motives span beyond the pursue of some ideal international order into a variety of other considerations such as deterrence, conflict preparation, or even the non-security externalities associated with alliances.

The second dimension that alliances may be thought to vary along is the function of capability aggregation in accomplishing the alliance's objectives.  Some alliances may have underlying objectives that are predicated upon the aggregation of resources, such as balancing alliances, while others may not and may emphasize alliances as non-security institutions, such as consolidating alliances \citep{morrow1991alliances, weitsman2004dangerous}.  In peacetime, alliances may become an institutional means of promoting policy coordination and accomplishing non-security objectives \citep{giberTaking2018}.  One interesting question is why states may use military alliances over other forms of international institutions when seeking to consolidate, expand, or pursue relationships.  Alliances are often the institution most frequently associated with high-level dialogue between between influential civil and military leaders, empowering them to accomplish these institutionalizing pursuits better than other forms of institutions, such as economic institutions \citep{pevehouse2005democracy, cranmerWorkingIndirect}.  
When these two dimensions are combined, much can be learned about alliance roles.

\begin{figure}[t!]
	\centering
		\includegraphics[height=5cm]{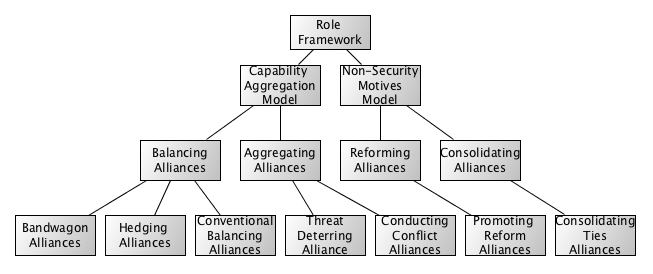}
	\caption{\textbf{Alliance Typology Tree.} Tree ascends in increasing order of abstraction.  All types of alliances belong to particular roles, which belong to particular models, which broadly capture all alliances.  }
	\label{type}
\end{figure}

The first role, Balancer, is reflective of conventional balance of power logic wherein capabilities are aggregated to achieve order-based goals.  
The second role, Aggregator, refers to a state that aggregates its capabilities by allying with others to counter a geopolitical threat and/or prepare for an inevitable conflict.  
These motives are agnostic to the broader international order.
The third, Reformer, is representative of a process where alliances are a means to promote internal reform within a state.  These reforms are promoted through institutional means, not necessarily security-based means, to pursue order-based motivations.  
The fourth and final role breaks from conventional wisdom, Consolidators may form alliances to consolidate economic or political ties through institutionalized arrangements.  Often these ties may be formed without paying close attention to broader international orders or capability aggregation.
It is worth noting that these four roles may not be the only roles adopted within the alliance network or the only type of alliances formed.  Alliances can be formed to serve a variety of purposes, as illustrated in Figure \ref{type} which shows the tree of alliance types and general levels of abstraction.
These roles, however, are consistent with general patterns that are expected to be the most prominent.

\begin{table}[]
\centering
\begin{tabular}{l|l|l|}
                                    & \textbf{Ordering} & \textbf{Narrowly Utilitarian} \\ \hline
\textbf{Capability Aggregating}     & Balancers         & Aggregators             \\ \hline
\textbf{Institutional} & Reformers         & Consolidators           \\ \hline
\end{tabular}
\caption{\textbf{Dimensions Sorting Alliance Roles.}  Alliances can be thought to vary according to whether they are ordering or narrowly utilitarian, and whether this is accomplished through capability aggregation or institutional mechanisms.}
\label{2by2}
\end{table}

It is worth clarifying what I mean when I refer to alliances within this conceptual framework.  Alliances, for the purposes outlined here, are defined in a way consistent with the ATOP project: ``Alliances are written agreements, signed by official representatives of at least two independent states, that include promises to aid a partner in the event of military conflict, to remain neutral in the event of conflict, to refrain from military conflict with one another, or to consult/cooperate in the event of international crises that create a potential for military conflict," \citep[238]{leeds2002alliance}.  Here, alliances are not defined with respect to their motive of aggregating power but rather as a treaty that contains provisions relating to security politics.  While these provisions may be important to the alliance, the alliance need not be formed solely for these provisions.  In fact, an institutionalist view of alliances may view such provisions as tertiary to the objective of forming a broader institution designed to accomplish other objectives.

\subsubsection{Balancer}
It is conventionally thought that a state's primary foreign policy objective is to ensure its security, often through countering the rise of a revisionist state or peer competitor, or preserving an ideal international order.  
Balancers aggregate resources to promote an ideal international order or preserve an existing international order.  
Capability aggregation occurs through the formation of alliances with like-minded states intended to maintain the balance of power  \citep{waltz1979theory, walt1990origins, mearsheimer2001tragedy}.  
The aforementioned example of the US forming NATO following World War II exemplifies a state adopting the role of Balancer \citep{holsti1970national}.  With a liberal order in mind, NATO was formed to aggregate capability to encircle the Soviet Union.  It may be argued that for states to be Balancers, they need not have common interests.  This is certainly true, in the case presented here, however, these states are aggregating capabilities for narrowly utilitarian objectives, and as such, may be thought of as Aggregators.  

The alliance behavior of both the state being balanced and the target of balancing, however, may often resemble one another.  
For example, during the Cold War the alliance behavior of both the US and the USSR are relatively similar.  In other words, even if a state is the target of balancing attempts, they too may adopt the role of Balancer to serve as a counterweight to the new balancing coalition.   

\begin{figure}[t!]
	\centering
	\begin{subfigure}[]{0.48\textwidth}
		\centering
		\includegraphics[height=5cm]{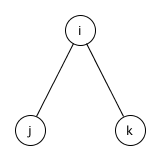}
		\caption{K-Stars or Preferential Attachment}
		\label{2star}
	\end{subfigure}
	\begin{subfigure}[]{0.48\textwidth}
		\centering
		\includegraphics[height=5cm]{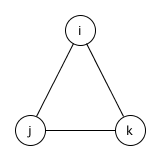}
		\caption{Triadic Closure}
		\label{triad}
	\end{subfigure}
	\caption{\textbf{Triadic Configurations.} K-stars may be reflective of a tendency towards preferential attachment, triangles measure triadic closure.}
	\label{triadconfig}
\end{figure}

The local alliance network of Balancers may have four observable features.  
Recent research has indicated that when constructing balancing coalitions, states create a larger coalition that includes many states as it lends legitimacy to their cause and the order they are attempting to preserve or establish \citep{keohane1974transgovernmental, adler2008spread, adler2009security, ikenberry2014power}.  This tendency can be measured through two network features.  First, the local alliance network of Balancers should show a tendency towards many ``high-degree" nodes.  Figure \ref{2star} presents the topographical representation of preferential attachment, often measured through $k$-stars wherein a node $i$ is connected to $j$ and $k$ but $j$ and $k$ are not connected.  
Second, these local alliance networks should be tightly knit and marked by a tendency towards triadic closure, illustrated in Figure \ref{triad}.  
This tight clustering would be an indication of similar preferences and a costly signal of the coalition's cohesiveness \citep{keohane1974transgovernmental, cranmer2012complex, cranmer2012toward}.  

Third, as Balancers may be forming alliances to uphold a particular order, it seems likely that they would form alliances with states of similar regime types.  States with similar regime types often have similar international preferences and are more likely to have similar foreign policy objectives \citep{gartzke1998kant, gartzke2000preferences}.  
The Holy Alliance following the Congress of Vienna seems to be a strong case of states of the same regime type having domestic incentives to uphold a conservative international order.  

Forth and finally, as states attempt to increase the strength of their balancing and order-preserving coalitions, or to aggregate resources, it may seem possible that stronger states may have to form alliances with weaker states, contrary to the expectations of the CAM \citep{morrow1991alliances}.  While there are increased autonomy costs for adding additional states \citep{morrow1991alliances, morrow2000alliances}, adding even weaker states will still marginally increase the overall size, strength, and perceived legitimacy of a coalition.  As such, one counter-intuitive expectation of this framework may be a tendency towards asymmetric alliances (with respect to capabilities) within the local alliance network of Balancers.  As previously mentioned, Balancing coalitions can also be engineered by revisionist states so it does not seem obvious that the local alliance network of Balancers should be truly marked by a tendency towards forming alliances with revisionist/non-revisionist states.  

\subsubsection{Aggregator}
An Aggregator attempts to aggregate capabilities to prepare for conflict, a motive that is agnostic to the broader international order.  
This role is fairly consistent with the logic of the CAM.  
Balancers form alliances as a means of pooling resources for creating or maintaining an international order, which may manifest through a desire to constrain a peer competitor and/or promoting a balance of power.  Aggregators, however, form alliances to pool resources to prepare for what is perceived to be an inevitable conflict.  Consider the case of Brazil, Argentina, and Uruguay in the Triple Alliance which marks a clear example of states coming together to prepare for a conflict, in this case, the Paraguayan War (1864-1870).  

Aggregators' alliance networks should be marked by several characteristics.  First and foremost,  Aggregators pay greater attention to the capabilities of their allies than Balancers.  Given that Aggregators form alliances based upon the expectation of an inevitable conflict, alliance-based autonomy costs and the security gains are much more salient \citep{morrow1991alliances, snyder1997alliance, morrow2000alliances}.  However, forming alliances with equally strong states may be difficult as there are constraints on available allies and sub-optimal alliances may exist.  As such, while it is expected that Aggregators do not form asymmetric alliances, they may not be able to form perfectly symmetric ones either.  
Second, if Aggregators believe that a conflict is inevitable, they may seek out many alliance partners and downplay autonomy costs.  As such, it might be expected to find a tendency similar to that of Balancers wherein the local network of Aggregators is constituted by many high-degree nodes.  
This is measured through $k$-stars or a degree term, illustrated through Figure \ref{2star}.  

Third, similar to Balancers, Aggregators may be more likely to form tightly knit alliance networks marked by a tendency towards triadic closure, illustrated in Figure \ref{triad}.  As \citet{cranmer2012toward} note, this may help bolster states' security through providing synergetic gains.  Fourth, and finally, states are said to make decisions about alliance reliability based upon attributes of potential partners \citep{crescenzi2012reliability}.  \citet{lai2000democracy} finds that democracies can more credibly commit their reliability as allies to one another.  In addition, autocracies may be more trusting of one another as they may have similar preferences \citep{gartzke2000preferences}.  
As such, regime type homophily is expected to be a common feature of an Aggregator's local alliance network.  As Aggregator's alliances may be fleeting alliances of convenience and include strange bedfellows, a tendency towards including revisionist or non-revisionist states does not seem likely.  

\subsubsection{Reformer}
The Reformer role is associated with states forming alliances as a means to push or be pushed by others to adopt internal reforms, likely economic or political liberalization, or influence policy.  The effect of alliances on political reforms or policy change is well documented \citep{pevehouse2005democracy, gibler2006alliances, giberTaking2018}, but typically considered as a positive externality as opposed to a chief motivation for alliance formation.  Consider the revised mission of the OAS following the Cold War.  Initially formed to expel Soviet influence in Latin America, the OAS was reorganized around a renewed set of principles, including a regionalist cooperative security order and respect for human rights, economic freedom, and liberal democracy \citep{pevehouse2005democracy, therien2012changing}. These new principles were consistent with the broader liberal internationalist order pursued by the West \citep{ikenberry2012liberal}.


Reformers are likely to have ego-networks that mirror the structure of Consolidators.  Similar to the prior alliance roles, Reformers should have ego-networks constituted by many high-degree nodes.  As the institutions designed to push reforms are typically very large and designed to form regional institutions, it seems inevitable that there would be many reform pushers and many reform pushees that are all interacting within a very large community.  Second, given that these institutions largely work through socialization \citep{pevehouse2005democracy}, a series of tightly knit alliance clusters seems to be essential to success.  If particular states were excluded from the community, then their socializing effects may not be as successful.  Third, Reformers' alliance networks are more likely to be constituted by many asymmetric alliances.  Within asymmetric dyads, stronger states may be able to exercise a greater degree of influence over weaker states in pushing their preferred policy outcomes, in this case, domestic reform \citep{morrow1991alliances}.  Fourth, initially, the alliance networks formed by Reformers should be constituted by regime heterophily, only to eventually be marked by regime homophily as reforms become successfully implemented.  When used as a vector for the diffusion of democracy, it is expected that reformers would be democracies forming alliances with autocratic reformees.  This dynamic is well documented, the US is often known to use alliances like the OAS to promote democratization abroad \citep{pevehouse2005democracy, therien2012changing}.

\subsubsection{Consolidator}
Finally, Consolidators form alliances to institutionalize existing economic or political ties or coordinate policy action.  This role has become more prevalent as it is increasingly common for states to use alliances to facilitate and institutionalize economic, political, or cultural exchange \citep{long2006trading, lake2009regional, fordham2010trade}.  These peacetime alliances may be be particularly useful in coordinating policy action \citep{giberTaking2018}.  In these cases the states may be agnostic to the broader order to capability aggregation and see the alliance as a useful institutional mechanism to consolidate a relationship \citep{powers2004regional}.  In the post-Napoleonic system, the German kingdoms, Prussia, and Austria formed alliances to pave way for the German Confederation and eventually, German Unification \citep{hartshorne1950franco, craig1995force}.  

Consolidators' ego-networks are likely to mirror those of Balancers and Aggregators with respect to network topology.  First, Consolidators are likely to have alliance networks constituted by many high-degree nodes.  Once states decide to adopt this role, they turn towards creating institutional arrangements with many states.  German Unification highlights this point clearly -- many kingdoms like Bavaria or Hanover formed many dyadic alliances with other kingdoms in an effort to bring many into the fold.  Second, there is likely to be a tendency towards triadic closure as these states seek to form tightly knit communities to assist consolidation.  In the case of German Unification, it would make little sense for some kingdoms to exclude others if the goal is to create a truly pan-Germanic community.  

Consolidators are likely to form alliances with states of disparate capabilities and regime types.  Consolidators' alliance ties may be more likely to emerge between states of disparate national capabilities as there is greater potential for heterophilous gains.  If two states have an interest in forming a firm relationship and seek heterophilous gains from one another, alliances may be an optimal institution allowing for such concessions and side payments \citep{morrow1991alliances, fordham2010trade}.  For example, a stronger state may seek trade concessions from a weaker state, but to give such concessions, the weaker state may demand defense.  These asymmetric alliances may provide a means to consolidate such ties.  Fourth, and finally, alliances may provide a means of bringing states of disparate regime types together.  Alliances are often considered a trust building mechanism \citep{kydd2001trust} and their effect in promoting preference conference is well known \citep{bearce2007intergovernmental}.

\section{Empirical Strategy}
This section introduces an empirical strategy for evaluating the previously introduced role-based framework.  It begins by introducing the ego-Temporal Exponential Random Graph Model (ego-TERGM), a model capable of identifying the previously discussed roles.  The following subsection then discusses details of each model estimated, including the time periods used, the number of roles estimated, and the covariates used.  It is worth noting that some states may enter analyses for certain periods, this occurs as states may form or dissolve, or not achieve the minimum number of alliances needed for any year within a period.  

\subsection{Inferring Alliance Roles Using the ego-TERGM}
To assess this role-based framework a novel statistical approach, the ego-Temporal Exponential Random Graph Model (ego-TERGM), is used.  This approach, outlined by \citet{campbell2017detecting}, decomposes a longitudinal network (such as the alliance treaty network) into its component ego-networks and clusters them according to similarity or difference across a variety of user-specified nodal, dyadic, temporal, or network variables.  In other words, if one views the interstate alliance network as constituted by a set of states and their allies, one may infer the roles states serve in the alliance network through examining patterns of how and why states form alliances.  

Previous work has indicated that the ego-TERGM, and its equivalent for cross-sectional networks, the ego-Exponential Random Graph Model (ego-ERGM), is useful for uncovering interesting properties of a network \citep{salter2015role, box2017role, campbell2017detecting}.  \citet{box2017role} used the ego-ERGM to examine the roles that interest groups adopt within the environmental lobbying coalition.  In examining a canonical social network, \citet{campbell2017detecting} uncovers a set of roles reflecting a unique aspect of labor negotiations and collective action in non-union shops.  


In simple terms, the ego-TERGM clusters a set of nodes and their local networks (ego-networks) within a broader longitudinal network into a user-defined number of time-invariant classes (clusters or roles) according to the similarity of these ego-networks (TERGM model parameters).  In more formal language, the ego-TERGM is a finite mixture model that attempts to assess mixtures of data generating processes (DGPs) for bootstrapped Temporal Exponential Random Graph Models (TERGMs) \citep{cranmer2010inferential, campbell2017detecting}.  This is accomplished through assigning each actor's longitudinal ego-network to a cluster according to the similarity in each network's TERGM parameters, which characterize the generating process for the network.  This model is an unsupervised and finite latent class model based upon the clustering of TERGM parameters, meaning that cluster memberships are determined entirely by estimated parameters.  For additional detail on the ego-TERGM, including its likelihood function, assumptions, estimation routine, and goodness of fit measures, the reader is referred to \citet{campbell2017detecting} and/or the SI Appendix.

The requirement that the analyst assign labels and meaning to each cluster is a limitation to any study that uses unsupervised machine learning to measure otherwise immeasurable phenomena.  Nevertheless, the alternatives to this relatively inductive approach are much worse.  Analysts either treat all alliances as capability aggregation by assumption, or they take the historian's approach of being inductive about every case such that there is no structure.  I mitigate the issue of inductive or \textit{ex post} label assignment through developing a theoretical framework that provides \textit{a priori} expectations about which roles or clusters should exist and which states should be assigned to these roles, and deductive observable implications that may be captured via ego-network structure.   In other words, while my approach may be inductive, I am principled and systematic in how I interpret raw historical data and cluster assignments, similar to how Bueno de Mesquita and co-authors use history in a systematic empirical basis \citep{de1985forecasting, bueno2011new}.  

In this model the user must specify a set of parameters.  This includes the minimum size an ego-network must achieve to be included in the analysis, the order of alters to include, the (T)ERG-based terms that roles are thought to sort on, and the number of roles that the model is allowed to fit.  For the models presented here, a state must have at least five allies to be included in the model.  This parameter value is chosen to increase the number of states considered for role assignments while also achieving a minimum size necessary to allow for model identifiability.  Only first order alters are examined as the allies of allies may not be thought to inform a state's role assignment.  In the following section the terms used to sort states are discussed.  Finally, in allowing the data to speak to the number of roles present, model BIC is allowed to inform the role assignment.  It is worth noting that in some cases theoretically-motivated constraints are imposed to prevent the model from selecting role cluster values that may be unacceptable, in this case, in excess of four total clusters of roles.  

\subsection{Data to Distinguish Alliance Roles}
This section discusses the model terms used to distinguish between role assignments, how they are measured, where they are sourced, and what constitutes the longitudinally observed network(s).  The analysis starts from a dyad-year dataset of all alliance treaties coded by the Alliance Treaty Obligations and Provisions dataset from 1816 to 2002 \citep{leeds2002alliance}.  These dyad-years are transformed into an undirected longitudinal network observed over 187 annually measured time steps.  A tie is present within this network when any two states are member to a treaty containing an offensive, defensive, neutrality, or non-aggression commitment.  The broader network of all treaties is preferred to the narrower network of particular commitments for two reasons.  First, for non-CAM roles, commitment may be independent of the role adopted.  For example, Consolidators and Reformers focus more upon the treaty and less upon the particular commitment when forming an institution.  Second, the presence of certain commitments over others may assist in distinguishing between particular roles.  While there are no strong \textit{a priori} reasons to suspect some roles would prioritize some commitments to others, including covariates for particular commitments allows the ego-TERGM to leverage additional information in producing role assignments.  

These networks are then partitioned into six historical time periods that are each analyzed distinctly.  These time periods are chosen to reflect the dynamic nature of state motives and the means through which distinct international systems or orders may influence the roles states adopt within the alliance network.
For a further discussion of these periods and the roles expected during them, the reader is referred to the SI Appendix. 
 
During each of these periods, I search the parameter space of models that are identifiable and produce the best fitting number of roles (or clusters) according to BIC while constraining the value to ensure parsimony.  When constraints are imposed they are done in a theoretically-motivated fashion to induce parsimonious assignments.  Table \ref{modelFitInformation} presents information on the models estimated on each of these time-periods.  

To distinguish between the roles adopted during this time I employ a variety of covariates that are used to distinguish each longitudinally observed ego-network.\footnote{The distributions of these covariates, their coding rules, and original data sources are discussed further in the SI Appendix.}  First, a simple edges term is used to distinguish between networks as one might expect roles to sort on the relative density of actors' ego-networks.  In other words, this model would be used to detect the number of dyadic alliances present in an ego-network.  Second, a measure for the number of $k$-stars within the network is used.  
Some roles may sort on $k$-stars as actors similar to Bismarck attempt to create dominant and expedient coalitions without becoming fully embedded, or attempt to form alliances with many different states.  Alternating $k$-stars is chosen to assist in model estimation. Third, a homophily term for regime type is used. This term captures the tendency of a state to form alliance with states of a similar or different regime type and can capture the formation of ideological communities.  
Data for this variable is taken from the Polity IV project \citep{marshall2002polity}, where a democracy is coded as any state with a regime score greater than 6 and a non-democracy is coded as any state with a regime score less than 7.\footnote{It is possible that states may seek out or seek to avoid autocratic regimes of a particular type when forming alliances \citep{weeks2008autocratic}.  Democracies may be more likely to push for domestic reforms in autocratic military regimes.  Accounting for variation in autocracy poses a tremendous modeling challenge as some autocratic types may not exist during some years, and as such, create model identifiability issues.  As such, a broader understanding of autocracy is adopted here.}  

In addition, two ``difference" variables are used to capture the tendency of two nodes to have disparate values on a variable.  The fourth term included measures the tendency for states to form alliances with either stronger or weaker states.  It is measured by the absolute difference in CINC scores for a dyad, with data sourced from the Correlates of War Project \citep{singer1972capability}.  
The fifth term included accounts for the absolute difference in binary indicators for whether a state is labeled as a revisionist state in a militarized dispute in that year.  
Data for this variable is also taken from the Correlates of War Project \citep{jones1996militarized}.  

Four attributes of the alliance are also used to distinguish between roles \citep{leeds2002alliance}.  The sixth, seventh, and eights terms included are whether an alliance formed contains a defensive, offensive, and/or secret commitment respectively.   Many alliance roles may sort on these covariates as states engineer and design their alliances to accomplish goals consistent with their role.  It is worth noting that when including these commitment-based covariates, the baseline category for treaties becomes the presence of a non-aggression or neutrality pact.  The last attribute is the number of years a dyad has had an alliance treaty.  This variable is designed to capture the temporal dependence that may exist between alliance years.\footnote{While the inclusion squared and cubed functions of alliance years may be ideal \citep{carter2010back}, their inclusion creates model identifiability problems.}  

It is worth noting that the choice of some covariates is constrained by whether there is variability for each dyad within an ego-network with respect to certain covariates.  For example, in some periods there is no variability in a state's ego-network with respect to offensive commitments.  
This invariability prevents the ego-TERGM from being identifiable for the broader network as initial parameter values cannot be estimated for all states.  For example, edge covariates for whether the alliance developed established institutions or was formed to counter common enemies were not included as they created identification problems.   The inclusion of a term for the number of triangles in a network also created problems as alliance ego-networks often reflect fully connected networks, and as such, also do not vary in dyadic change statistics.\footnote{It should be noted that the terms discussed here and used for the ego-TERGM differ from those later used to detect commonalities in ego-network structure among states of the same role.}   

\begin{table}[]
\centering
\scalebox{0.8}{
\begin{tabular}{llll}
\hline
 \textbf{System or Period}                               & \textbf{Time Span} & \textbf{Num. Roles Fit}  & \textbf{Covariates Used}    \\
 \hline                                                                                                                                                                                                       
\rowcolor{Gray} Congress of Vienna           & 1816-1848 & 3  & \begin{tabular}[c]{@{}l@{}}Edges, Alternating K-Stars (0.5), Regime Homophily, \\ CINC Difference, Revisionist Difference, \\ Defensive Commitments, Alliance Years\end{tabular}                                           \\
Nationalism and Bismarckian  & 1849-1890 & 4  & \begin{tabular}[c]{@{}l@{}}Edges, Alternating K-Stars (0.5), Regime Homophily, \\ CINC Difference, Revisionist Difference,\\ Defensive Commitments, Alliance Years\end{tabular}                                            \\
\rowcolor{Gray} Pre-WW1                      & 1891-1918 & 2  & \begin{tabular}[c]{@{}l@{}}Edges, Alternating K-Stars (0.5), Regime Homophily,\\ CINC Difference, Revisionist Difference,\\ Defensive Commitments, Offensive Commitments,\\ Secret Provisions, Alliance Years\end{tabular} \\
Interwar                     & 1919-1945 & 1 & \begin{tabular}[c]{@{}l@{}}Edges, Alternating K-Stars (0.5), Regime Homophily,\\ CINC Difference, Revisionist Difference,\\ Defensive Commitments, Offensive Commitments,\\ Alliance Years\end{tabular}                    \\
\rowcolor{Gray} Containment and Bipolar      & 1946-1991 & 4  & \begin{tabular}[c]{@{}l@{}}Edges, Alternating K-Stars (0.5), Regime Homophily,\\ CINC Difference, Revisionist Difference,\\ Alliance Years\end{tabular}                                                                    \\
Liberal International        & 1992-2002 & 4  & \begin{tabular}[c]{@{}l@{}}Edges, Alternating K-Stars (0.5), Regime Homophily,\\ CINC Difference, Revisionist Difference,\\ Alliance Years\end{tabular}      \\
\hline                                             
\end{tabular}
}
\caption{\textbf{Model Fit Details:} Each row represents the model estimated and presented for each time period and the information relevant to its estimation.}
\label{modelFitInformation}
\end{table}
	
	
	
	
	
\section{Results and Discussion}
The results for my analyses uncover a role-structure consistent with the previously introduced role framework of alliances.\footnote{Models were validated using a combination of within-sample validation based on BIC for the number of roles fit parameter and external validation based upon historical expectation.  Given that labels and roles are not known \textit{a priori} cross validation of role assignments is not possible.  A broader discussion of model validation is provided in the SI Appendix.}  Alliances, it appears, are formed for a variety of purposes beyond those expected by the Capability Aggregation Model.  In particular, across time states form alliances to adopt non-security aggregating roles, including Consolidator and Reformer.
The results for each period are discussed in turn.\footnote{When discussing the roles states adopt within each period, it is important to note that these results are based upon general patterns of behavior consistent across all states included.  There may occasionally be country-to-country differences in how these roles are performed, which may be nuances too subtle to be teased out.} 
This section will discuss each period and the roles uncovered, in turn, and conclude by discussing the generative structure for each role. 

\subsection{Period Roles}
\subsubsection{Vienna System Roles}
In the years following the Napoleonic Wars, the victors established a principled order designed to promote stability, prevent conflict, and uphold autocratic rule \citep{taylor1954struggle, jervis1985balance, schroeder1996transformation}.  The German kingdoms also moved to promote unification through consolidating relationships and creating a pan-Germanic confederation \citep[599-606]{schroeder1996transformation}.  While the order was designed to prevent conflict, force was occasionally seen as necessary to prolong the post-Napoleonic order \citep[736-740]{schroeder1996transformation}.  As such, states would be expected to adopt the role of Balancer, Consolidator, or occasionally, Aggregator.  

\begin{figure}[t!]
	\centering
		\includegraphics[height=15cm]{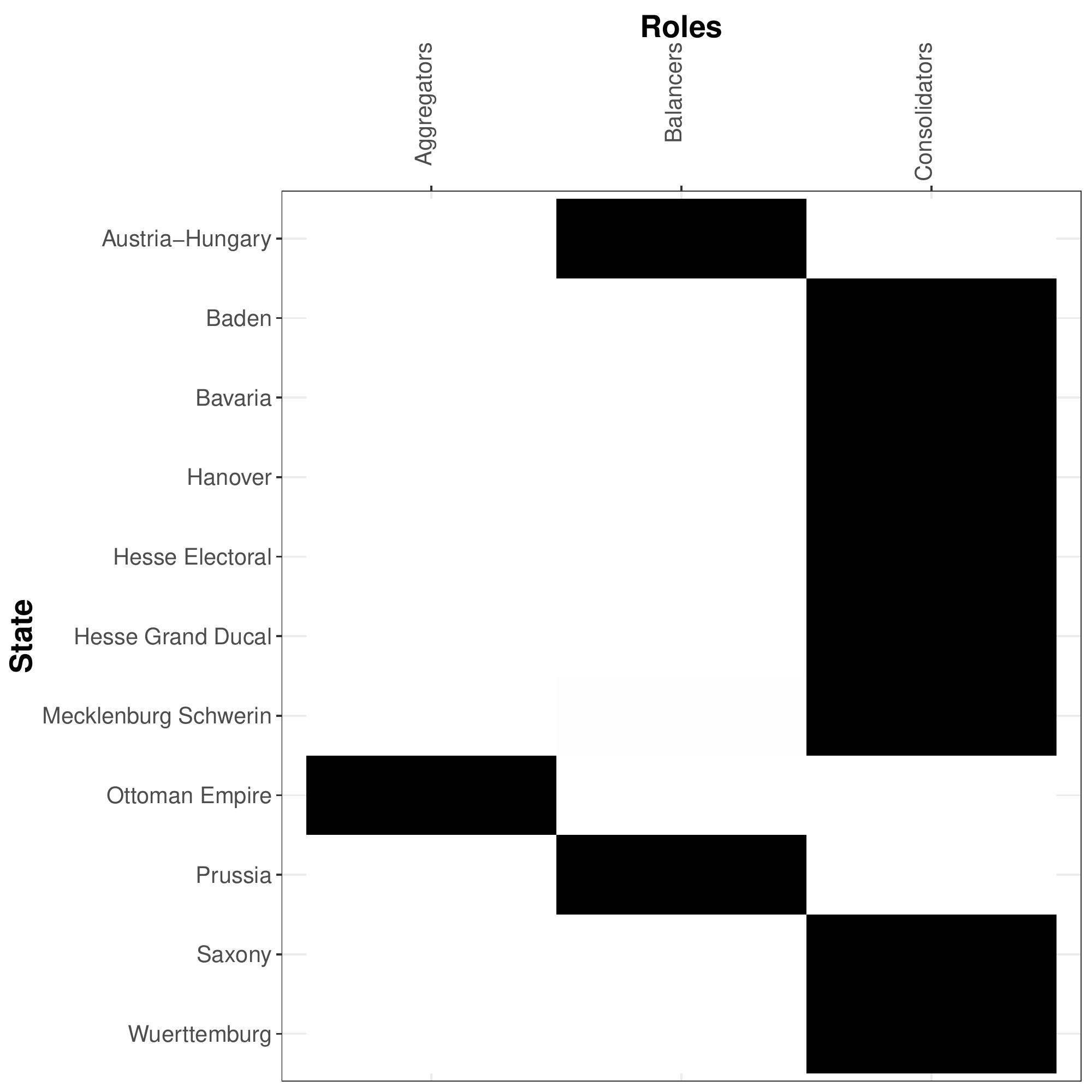}
		\caption{\textbf{Ego-TERGM Role Assignment Probabilities by Country, Vienna System.}  Darker cells show higher probabilities of assignment to a particular role.  Country labels refer to their Correlates of War System Membership Data name which may include legacy names for latter periods.}  
		\label{Vienna}
\end{figure}

During the Vienna System, the period spanning 1816 to 1848, a role-structure consisting of three roles, Consolidators, Balancers, and Aggregators, is uncovered and found to produce superior model fit according to BIC. Table \ref{modelFitInformation} presents the parameters used for this (and all other) ego-TERGM(s) fit.  
Overall, the \textit{a priori} expectation is confirmed: states appear to either adopt the role of Consolidator, Balancer, or a variation of Aggregator.  Results for this period are presented in Figure \ref{Vienna}.\footnote{To understand the relative contribution of each model term, a sensitivity-based approach is utilized.  This routine and its results for each period are discussed in the Si Appendix.}  

First, one set of roles were adopted by the smaller kingdoms that were member to the German Confederation.  The German Confederation, made up of kingdoms like Bavaria and Hanover, as well as powers like Austria and Prussia, was a means of promoting economic and regional stability \citep{schroeder1996transformation}.\footnote{Note that the names discussed here are their historical names while the names presented in in the role assignment figures contain their Correlates of War System Membership Data legacy names.  For example, Correlates of War labels Prussia as its later legacy Germany, and labels the Ottoman Empire as its later legacy Turkey.  This naming is preserved to ensure consistency throughout the manuscript.}    The model's selection of these states to include in the same cluster, and not the powerful German Confederation members like Prussia or Austria, is indication that these states may have formed a series of alliances to consolidate and bring order to existing economic and political ties.  This is consistent with the previously discussed Consolidator role and the expectation of diplomatic history \citep{schroeder1996transformation}.  

Second, the powers of the German Confederation, Austria and Prussia, are found to adopt a role distinct from their German Confederation co-members that closely resembles the Balancers role.  These two states are partners to the well known Holy Alliance (and later the Quadruple and Quintuple Alliances) formed to maintain the Vienna System and its set of norms.  Initially this alliance was formed between Austria, Prussia, and Russia to make the world safe for monarchs, but was expanded to include the United Kingdom and France as to maintain norms against violence and ensure an adequate balance of power by creating a dominant order-preserving coalition \citep{taylor1954struggle, jervis1985balance, schroeder1996transformation, bridge2014great}.  This is in contrast to the expectations of \citet{schroeder1992did,schroeder1996transformation} who argued that the Vienna System did not rest upon a balance of power, and that during this time the powers did not view power aggregation as essential to sustaining the system.  France, Great Britain, and Russia are not included in the analysis for this period as they never ally with at least five other states.  This logic, and the motivation for the alliances formed by Austria and Prussia at this time, is consistent with the Balancers role discussed in Section \ref{theory}.  

The third and final role identified is solely constituted by the Ottoman Empire (modernly, Turkey).  During the Vienna System, the Ottoman Empire found itself party to a multi-state alliance for several months in 1840.  This alliance alliance was formed to ensure Egypt would accept a peaceful negotiated settlement to the Second Egyptian-Ottoman War (1839-1841).  This case reflects a particularly odd dynamic rarely considered -- alliances formed as a means of promoting dispute resolution through resource aggregation.  This alliance was not formed prior to the conduct of conflict and as such does not cleanly match the logic of the Aggregators role.  Instead it represents a variation on this logic, resources aggregated after conflict to incentivize a winning party's acceptance of a negotiated settlement, as opposed to continue executing the conflict for additional gains.   This version of the Aggregators role, however, is consistent with historical intuition per \citet[736-740]{schroeder1996transformation} as the Ottoman multi-state alliance was ultimately formed in an effort to prevent the collapse of the Vienna System.

It is worth noting that Russia, France, and the United Kingdom are conspicuously omitted from my analyses.  The reasons are a result of model identifiability -- during no period do the local alliance networks of any of these countries contain enough states to allow a model to be estimated.  For example, during this time the UK typically only has three alliances, having four alliances with Prussia, Austria-Hungary, Russia, and the Ottoman Empire in 1840.  An ego-network of this size often provides insufficient variation or evidence to allow for a model to be estimated.   This is not ideal from a historical perspective, but it is not obvious what alternative would be suitable and allow for these states' inclusion.

  \subsubsection{Bismarckian System Roles}
\begin{figure}[t!]
	\centering
		\includegraphics[height=15cm]{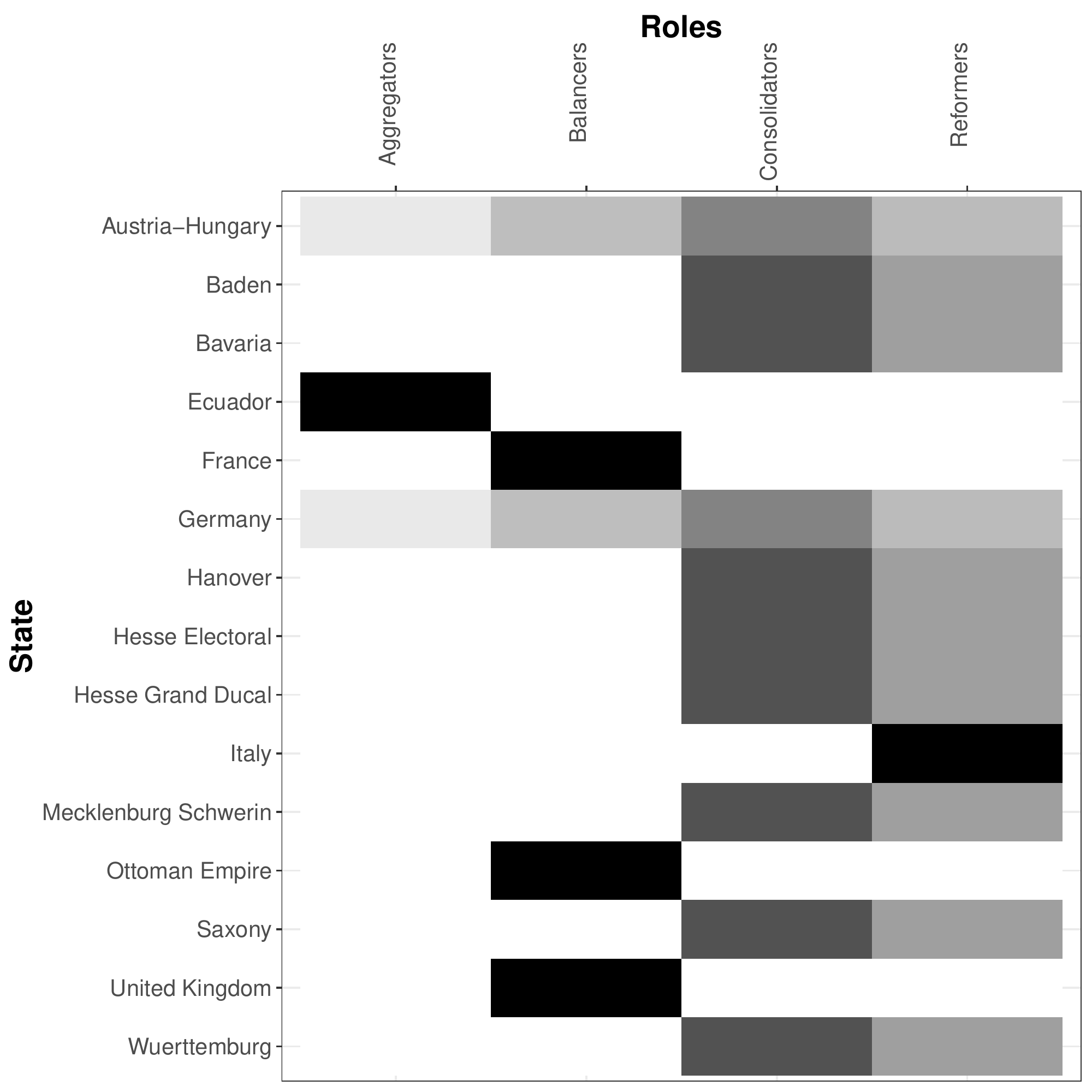}
		\caption{\textbf{Ego-TERGM Role Assignment Probabilities by Country, Bismarckian System.}}  
		\label{Bismarckian}
\end{figure}

The Bismarckian System, which spans 1849 to 1890, begins as a series of revolutions spread across Europe.  These revolutions make it likely that Reformers will emerge in an effort to nudge states either towards or away from democracy \citep{craig1995force}.  In addition, many conflicts, most notably the Austro-Prussian War, occur that might lead states to adopt the Aggregators role \citep{craig1995force}.  Under Bismarck, German unification escalates, the German kingdoms would likely continue to adopt the role of Consolidator as they attempt to transition to a Unified Germany with Bismarck at the helm \citep{hartshorne1950franco}.  During the Bismarckian System, four distinct roles are detected and $G$ and found to produce the best fitting model according to BIC.   The role assignments for this period, presented in Figure \ref{Bismarckian}, demonstrate a great degree of variation that is mostly consistent with theoretical and historical intuition. 

Similar to the prior period, many German kingdoms find themselves clustered in the same role with a relatively high degree of certainty.  The inclusion of Prussia and Austria within this role alongside many smaller German kingdoms indicates two significant legacies of Bismarck's foreign policy.  During this time, Bismarck used many of the alliances with the German kingdoms as a means of consolidating existing relationships in an effort to further promote German Unification \citep{hartshorne1950franco}.  However, it is quite clear that Bismarck also saw his chief role as a Balancer and through a thoughtfully engineered alliance system which included Italy, Austria-Hungary, and Russia, he sought to promote stability \citep{craig1995force, snyder1997alliance}.  As such Prussia and Austria appear to be mixture of Consolidators and Balancers.  

On the outside of Bismarck's alliance system often laid France, whom this League of the Three Emperors with Austria and Russia was designed to exclude.  Intuitively then France is in a separate role with the United Kingdom and Ottoman Empire.  This clustering make sense as the Ottoman Empire and England often found themselves at odds with the League of the Three Emperors \citep{taylor1954struggle, snyder1997alliance}.  These states, on the outside of Bismarck's core, do not appear to reflect any of the roles discussed in Section \ref{theory} and appear to be on the outside of a core-periphery system.  As a counter-weight to Bismarck's order-preserving and balancing coalition, France, the Ottoman Empire, and England too appeared to have adopted the Balancers role.  

The third role is solely constituted by Ecuador with very low-level probabilities of assignment from Germany and Austria.  As there are no other countries in Latin America, little alone the Western Hemisphere, who have at least five alliances at this time, they appear to be the most active state with respect to alliance politics.  During this period Ecuador collaborated with many other countries, including Peru, Chile, and Bolivia against Spain in the Chincha Islands War.  They are also engaged in three other conflicts during this period, which seems to indicate that they adopted a role most consistent with the Aggregators role as they seemed to be constantly engaged in conflict and as such, behooved to find wartime partners.  The low-level probabilities of assignment for Austria and Prussia also make sense as for a brief time they prepared for and engaged in the Austro-Prussian war.  

The final role reflects the Reformers role, and is constituted largely by Italy with all Consolidators having a low-level probability of assignment.  Following 1848, revolution spread across Europe, dramatically changing its geopolitical landscape \citep{taylor1954struggle, schroeder1996transformation, bridge2014great}.  In particular, these revolutions shaped the foreign policy trajectories of Italy, Austria, and Germany.  In particular, the revolts in Italy had geopolitical implications that ultimately unraveled the European Order and brought France and Austria to battle over Italian independence in 1859 and Italy and Austria to fight in 1866.  With each conflict, Austria's hold on Italy weakened and a path for Italian unification was paved.  As Italy became increasingly independent, France and England used their clout and alliance ties with Italy to push for Italian independence and democratization.  As such, Italy can be described as a Reformer.

\subsubsection{Pre-World War I System Roles}
\begin{figure}[t!]
	\centering
		\includegraphics[height=15cm]{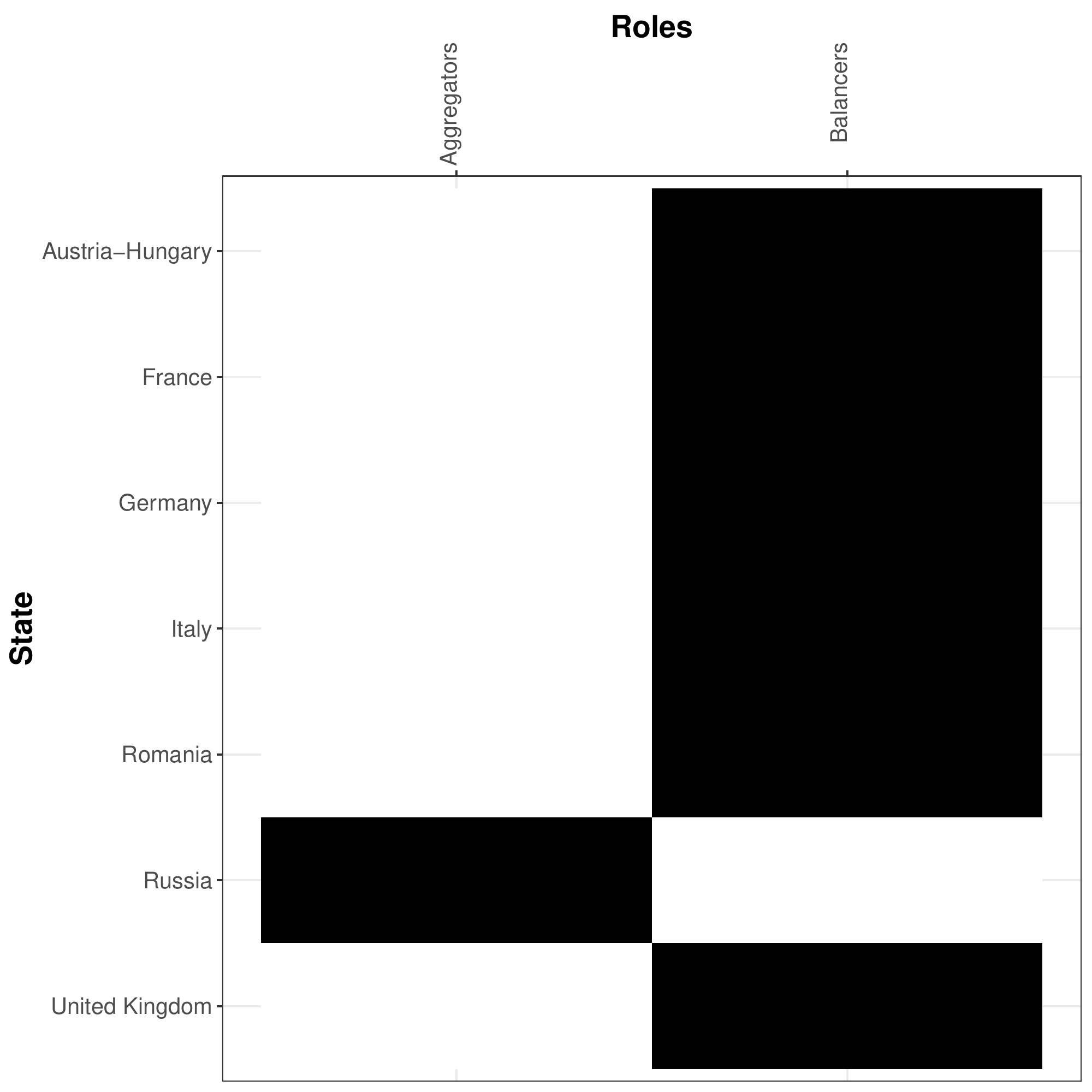}
		\caption{\textbf{Ego-TERGM Role Assignment Probabilities by Country, Pre-World War 1 System.}}  
		\label{preww1}
\end{figure}

The Pre-World War 1 period, spanning 1891 to 1918, represents a tumultuous time wherein Bismarck's sophisticated alliance system was handed to a set of diplomats that many considered to be professionally incompetent \citep{taylor1954struggle, kissinger1994diplomacy}. During this period the lasting effects of Bismarck's \textit{realpolitik} system are felt and as competition between states increased, there were increased pressures for states to arm and aggregate power.  This system appears to be a much simpler role system than the prior periods, as two role are detected and found to produce optimal model fit according to BIC.  
Results presented in Figure \ref{preww1} show support for historical expectation as Balancers and Aggregators are both dominant.  Overall, the the theoretical intuition is corroborated to some degree, although the prediction that Aggregators would be more prevalent than Balancers is not confirmed.

The first role is solely composed of Russia, who during the lead up to World War I would go on to be a member of the Triple Entente with Great Britain and France.  Russia was particularly unique from its Triple Entente counterparts, however, in that they increasingly expansionist and turned their eye on the Far East \citep{kissinger1994diplomacy}.   Nicholas II was of the opinion, among many others, that being a great power required territorial expansion \citep{kissinger1994diplomacy}.  This expansion made it clear that Russia was accruing alliances for the purpose of aggregating its power and expanding its influence.  As such, Russia appears to have adopted the Aggregators role. 

The second role is composed by the remaining states who have at least five allies during this time: United Kingdom, France, Germany, Austria-Hungary, Italy, and Romania.  Many of the states adopting this role are members of either the Triple Entente or the Triple Alliance which are alliances thought to be formed to balance other great powers \citep{taylor1954struggle}.  The Triple Alliance, for example, was designed to balance against France and to prevent attack on Italy or Germany.  The Triple Entente was mostly formed as a counterweight to the Triple Alliance.  As such, I would refer to these states as Balancers.

\subsubsection{Interwar System Roles}
\begin{figure}[t!]
	\centering
		\includegraphics[height=9cm]{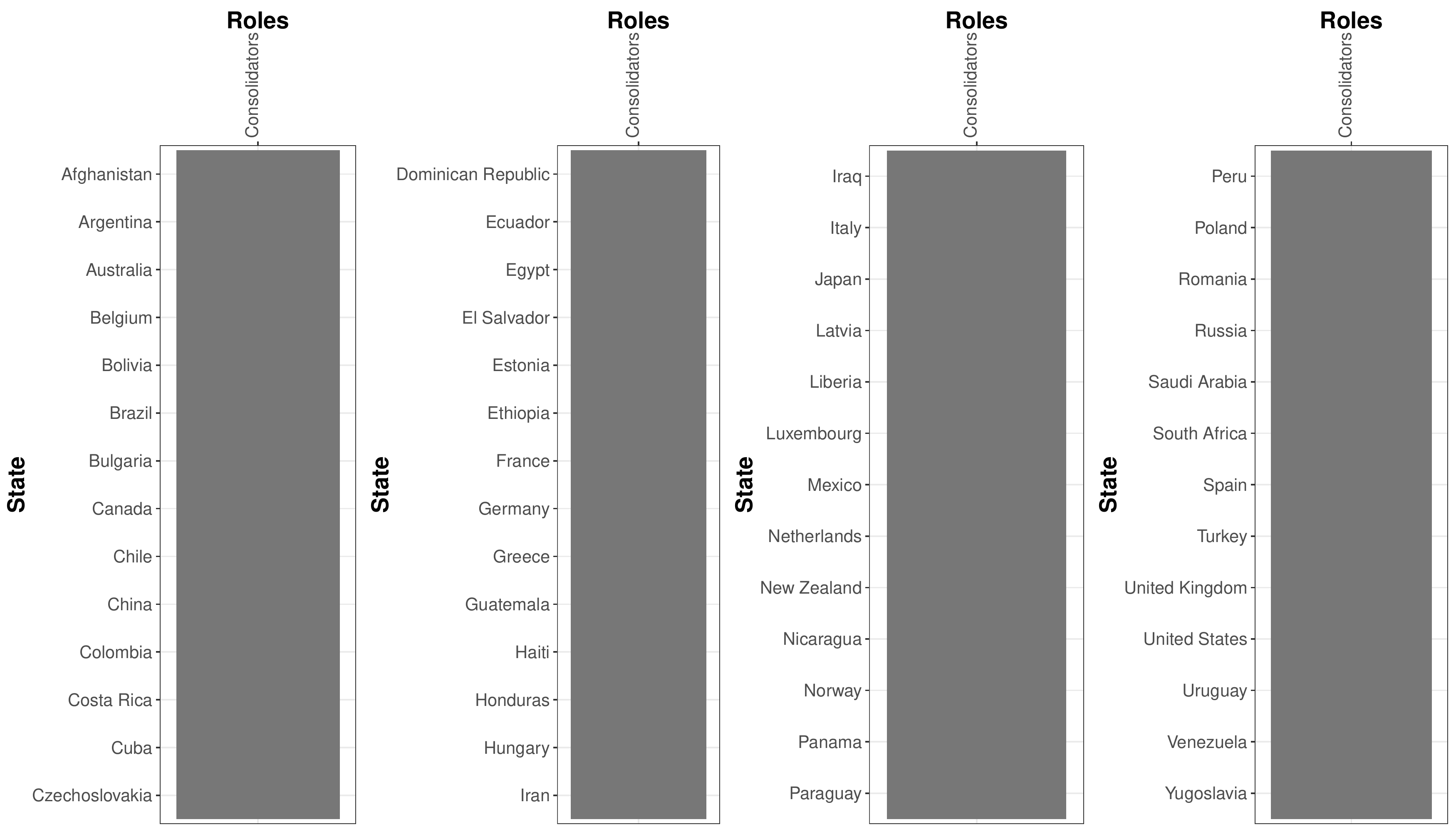}
		\caption{\textbf{Ego-TERGM Role Assignment Probabilities by Country, Interwar System.}}  
		\label{interwar}
\end{figure}

After World War I, the role structure of world politics became much more complicated \citep{carr1939twenty, siverson1982alliances, kissinger1994diplomacy}.  While attempts were made to consolidate relationships through the failed League of Nations or to promote liberal reforms, there are were many salient cases of state competition \citep{carr1939twenty}.  As such, it is unclear what the role structuring principles of this period were.  Given this expected heterogeneity \citep{siverson1982alliances}, it is interesting that a homogenous role system consisting of only one role is found to produce superior model fit.  Results from the model are presented in Figure \ref{interwar}.  While theoretical expectation would expect to find some combination of roles, only Consolidators were detected

The estimation of any model with $G > 1$ produces a more poorly fitting model when accounting for the penalization of additional roles through BIC.  This indicates that the ego-network for all states included in the model appear similar, at least with respect to the covariates specified.  During this time the post-WW1 order was created and states began to focus upon the consolidation of political and economic relationships to maintain the broader international order established in Versailles \citep{carr1939twenty, kissinger1994diplomacy, craig1995force}.  This is not precisely consistent with theoretical logic, as it may be expected that states also form alliances in an effort to aggregate power to counter a rising Germany or to prepare for conflict  \citep{siverson1982alliances, kissinger1994diplomacy}. 

\subsubsection{Bipolar System Roles}
\begin{figure}[t!]
	\centering
		\includegraphics[height=12cm]{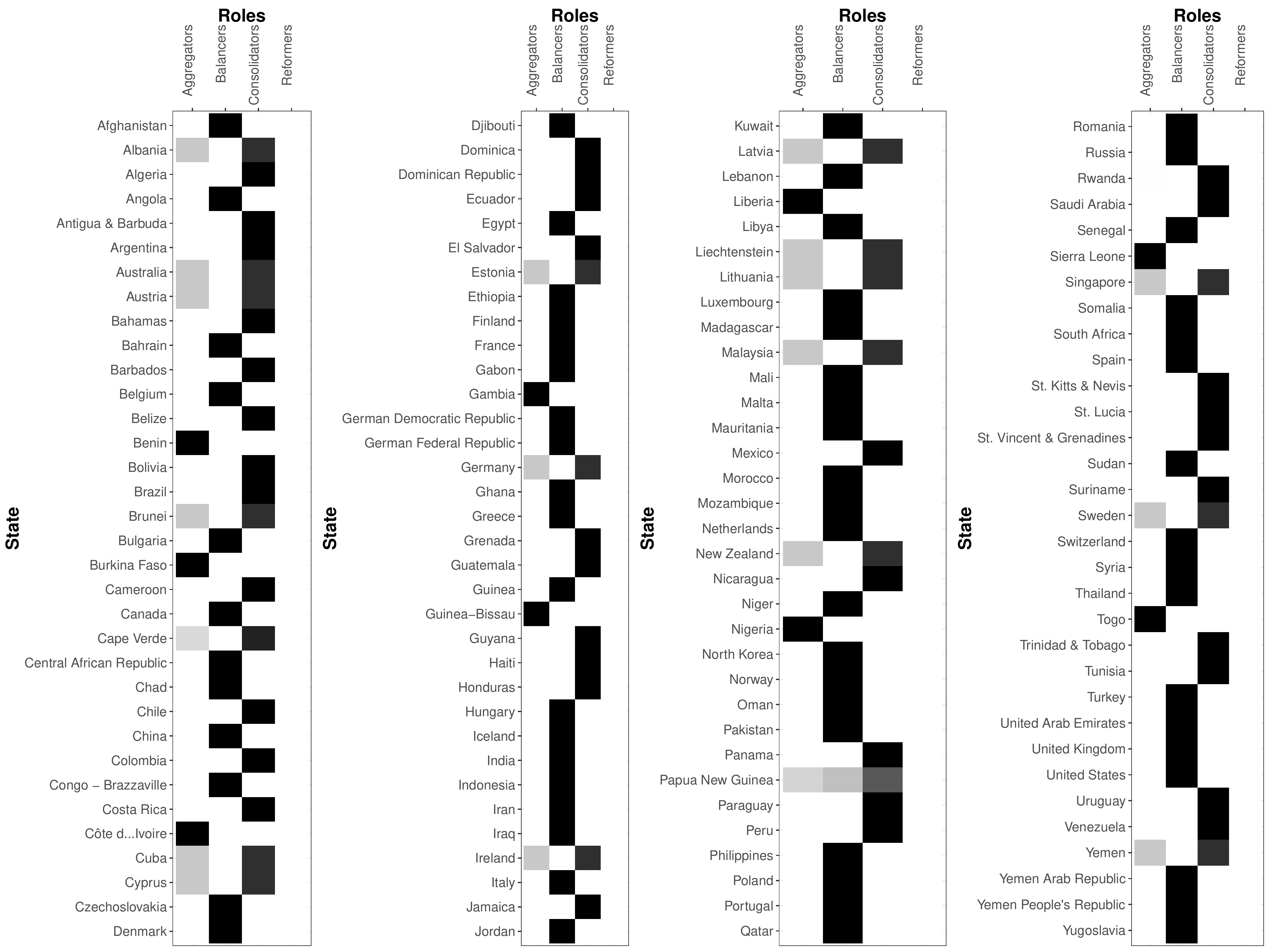}
		\caption{\textbf{Ego-TERGM Role Assignment Probabilities by Country, Bipolar System.}}  
		\label{containment}
\end{figure}

This bipolar system emerges in 1946 and lasts until the dissolution of the Soviet Union in 1991.  Following the resolution of World War II international politics began to organize along two distinct alliance blocs: NATO and the Warsaw Pact.   With that being said, however, it does not seem clear that states within each of these blocs form alliances for distinct reasons as they are members of coalitions balancing each other.  As previously discussed, this period should be dominated by Aggregators, Balancers, and Consolidators.  Four roles are detected and found to produce  the best fitting model according to BIC.   Of these roles, 133 different states are assigned to three different roles.  Intriguingly, in addition to the roles expected, adding an additional role is found to improve model fit.  However, no state has a probability of assignment to this role greater than 0.00001.  We refer to this as the Reformers role, which while likely to be rare at this point, may still exist in generality.   This model is estimated according to the parameters discussed in Table \ref{modelFitInformation}.  We will discuss these four roles, highlighting particularly powerful states that adopt these roles.  Results from this model are presented in Figure \ref{containment} and largely consistent with prior expectation.  

The first role contains the most number of states and is reflective of the Balancers role.  Among these members are prominent early NATO countries, including the United States, the United Kingdom, Canada, Belgium, and France, but also late comers such as Turkey and Spain.  Additionally, all Warsaw Pact countries excluding Albania are included in this role.  During this period these two alliance blocs commonly formed alliances as a means of containing and balancing the influence of the other \citep{waltz1993emerging, kissinger1994diplomacy}.  

The second role is predominantly constituted by smaller states that do not directly participate in either of these alliance blocs (except Albania).  Many of these countries are early and late coming members of the OAS, such as Brazil, Argentina, and St. Lucia, while others are members of the Arab League, such as Algeria or Tunisia.  During this time many of these alliances were formed to consolidate existing ties through promoting coordination and cooperation \citep{lebow1994long}.   These motives are consistent with the Consolidators role as states formed alliances to institutionalize existing ties.  While the United States may have been using the OAS as an opportunity to balance Soviet influence and promote western-style modernization, it appear that many of its OAS allies may have had different motives, viewing the alliance as an opportunity to consolidate their relationship with each other and with the hegemon \citep{weeks1991beyond, mikoyan2018russia}.  This may be useful as a regime's support from the United States may have predicted the United States' support for the regime and thus its survival. \citep{slater2008geopolitics, stodden2016interests}  In such cases, participation within the OAS may have been useful for consolidating economic and political relationships, producing a series of positive externalities, but such participation may not have been totally voluntary \citep{weeks1991beyond, slater2008geopolitics, stodden2016interests}.\footnote{There has been a great amount of highlighting that many of the asymmetric relationships during this time were coercive and marred by the use of colonial and imperial power \citep{slater2008geopolitics}.  As such, one might view the hegemon as exercising power over the subaltern to ensure that the subaltern participates in alliances in an effort to consolidate the asymmetric relationship and bolster the perceived legitimacy of the American-promoted liberal internationalist order \citep{slater2008geopolitics}.}    

The last role discovered, adopted by Benin, Burkina Faso, Gambia, Guinea-Bissau, Ivory Coast,  Liberia, Nigeria, Sierra Leone, and Togo, appears to be a variation of the Aggregators role wherein states form a larger institution to aggregate power in an effort to bring order to a region and respond to internal threats.   Many of these states are party to the Defense Pact of the African and Malagasy Union or the Economic Community of West African States (ECOWAS), both institutions designed to uphold norms against territorial aggression and military confrontation in Africa through aggregating capabilities and deterring armed conflict \citep{ukeje2005economic}.

\subsubsection{Liberal International System Roles}
\begin{figure}[t!]
	\centering
		\includegraphics[height=12cm]{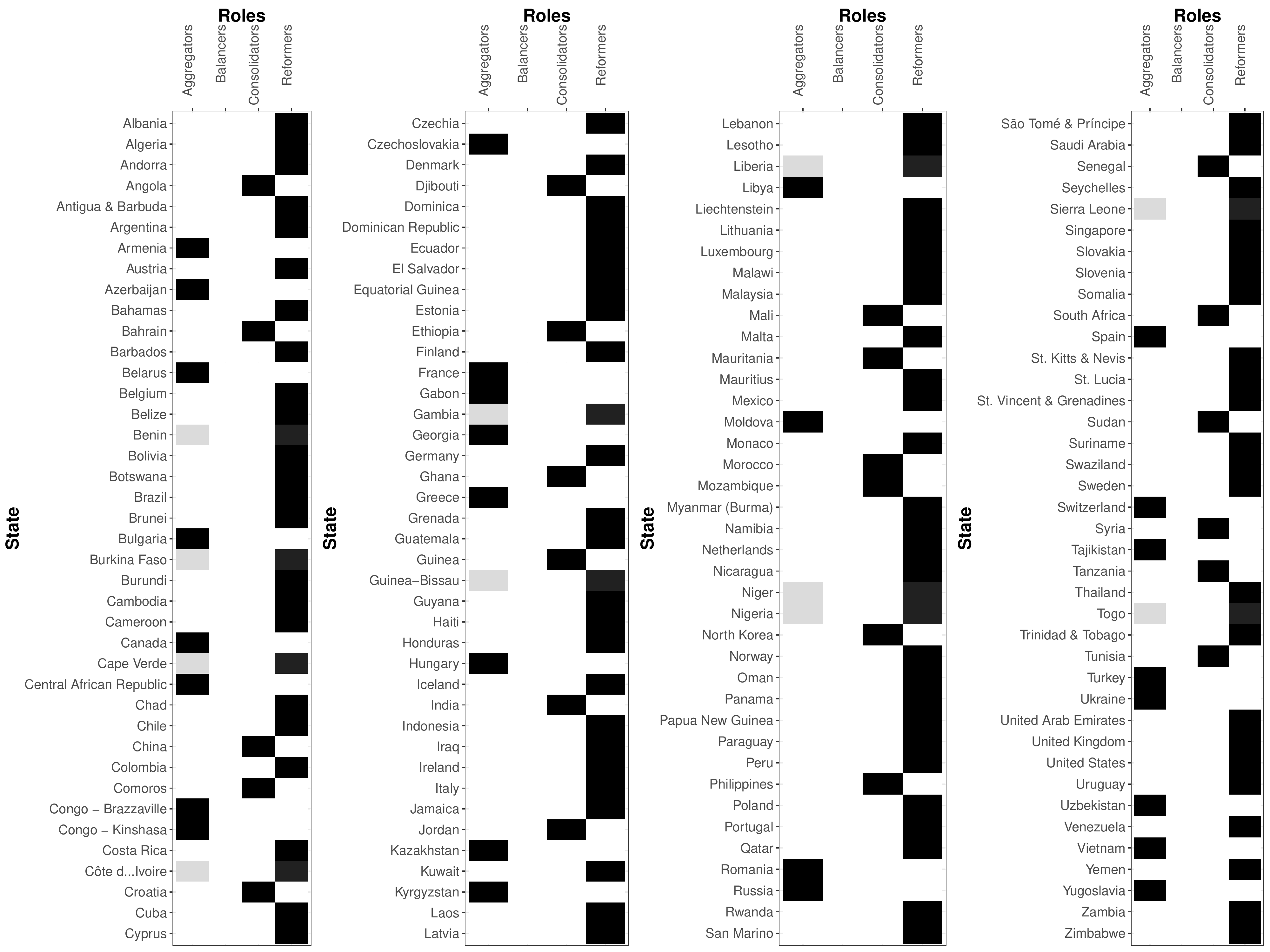}
		\caption{\textbf{Ego-TERGM Role Assignment Probabilities by Country, Liberal International System.}}  
		\label{unipolarity}
\end{figure}

The final role system considered is that of the Liberal International System that emerges following the dissolution of the Soviet Union in 1991 and is considered until 2002, the last year collected in ATOP.  During this time the US has its ``unipolar moment" which radically transformed international politics and created a new order based upon liberal international norms \citep{ikenberry2012liberal, ikenberry2014power}.  New constraints on competitive behavior has lead states to use alliances to accomplish nonconventional objectives consistent with the Reformer or Consolidator roles \citep{pevehouse2005democracy}.   During this period a model is estimated according to the parameters in Table \ref{modelFitInformation}.  A role structure consisting of four roles is detected and found to produce optimal model fit according to BIC.  As is with the prior case, 158 countries are assigned to three different roles even though four roles are fit. Results from the ego-TERGM fit on this period are presented in Figure \ref{unipolarity}, which present results that are largely in line with theoretical expectation.  

The first role is constituted by the most countries and is reflective of a process where many states view military alliances as an opportunity to promote some broader order through institution building.  This is representative of the Reformers role.  This primarily includes countries within the OAS which are party to its renewed mission and effort to promote a regional security community based upon respect for democracy, free trade, and cooperative security \citep{therien2012changing}.  During this time, many of the democracies within the OAS pushing autocracies to liberalize through the Unit for the Promotion of Democracy (UPD), such as the democratic United States and Mexico tied to autocracies like Venezuela, Guatemala, Haiti, and Nicaragua \citep{pevehouse2005democracy, slater2008geopolitics, therien2012changing}.  Parallel narratives also emerge within NATO and ECOWAS, wherein former security institutions adopt a renewed mission of assisting states in pursuing reforms and promoting a regional security community or order \citep{adler2008spread}.  

The second role is constituted by many other peripheral states that are active outside of NATO (while including some NATO member countries).  Many of these states, like France and Russia, have at times resisted the liberal international order promoted by the United States.  As such, it does not seem clear that these states adopt the Reformers role, and at the same time, do not appear to cleanly fit into any other roles.  Many of these states may be more reflective of an unexpected Aggregators role as many participate in conflict during this time.    

The third and final role closely resembles the Consolidators role and is constituted by a set of African, Asian and Middle Eastern countries who have formed alliance-based institutions like the Arab League or the Shanghai Cooperation Organization.  These institutions are largely means to prevent aggression and provide an institutional means to consolidate existing economic, political, or cultural relationships and promote cooperation.  

\subsection{Assessing the Role Generative Structure}
To assess the generative structure of the roles uncovered by the ego-TERGM, four bootstrapped pseudolikelihood TERGMs are fit on all ego-networks associated with each role for each systemic period.  This routine, introduced by \citet{campbell2017detecting}, allows analysts to understand the network features that are prevalent or not for all networks associated with a role.\footnote{For additional discussion of the TERGM estimated through bootstrapped pseudolikelihood, I refer the reader to \citet{cranmer2010inferential}.}  The models presented here are the best fitting identifiable models associated with each role.\footnote{In the SI Appendix, TERGMs fit on each role for each time period are introduced and discussed.  For additional detail on the covariates used for these TERGMs that are not used for the ego-TERGM, the reader is referred to the SI Appendix.}  

Results for this routine are presented in Table \ref{roleTERGMs}.\footnote{Goodness of fit diagnostics for these models are presented in the SI Appendix.}   Overall, the results illustrate that the observed role-generating processes pretty closely mirrors the expected role-generating processes.   
The pooled TERGM fit on all Balancer ego-networks finds a tendency towards high-degree nodes, marked by a positive and robust effect for GW Degree, and triadic closure, found by a positive and robust effect for the Triangles term.  This makes sense as Balancers may create large and tightly knit coalitions to legitimize their ideal international order  \citep{adler2008spread, adler2009security, ikenberry2014power}.  In addition, the model also discovers a tendency towards regime homophily and capability heterophily, marked by positive and robust effects for both the Regime Homophily and CINC Difference terms respectively.  Balancers may be more likely to bring in weaker states to further illustrate the breadth of support for their ideal order, but this desire for a breadth of support may be constrained by foreign policy preferences that stem from regime type \citep{gartzke1998kant, gartzke2000preferences}.  

The Aggregators model also uncovers results consistent with theoretical expectation.  Aggregators' ego-networks have a tendency towards both triadic closure and high-degree nodes, indicating fairly large and tightly knit alliance clusters that would produce synergetic gains to state security \citep{cranmer2012toward, cranmer2012complex}.  In addition, it also appears that Aggregators are more likely to form alliances with states of similar regime types.  This makes sense as Aggregators must form alliances with states they can trust given the potential security costs of ally renegement \citep{lai2000democracy, gartzke2000preferences}.   Finally, Aggregators do not illustrate a strong tendency to form symmetric alliances over asymmetric alliances.  Theoretical expectation would hold that Aggregators would be more likely to seek allies with similar capabilities \citep{morrow1991alliances}.  However, as noted, a null effect may also be expected as alliance choices are constrained by the allies willing to collaborate with the state in an inevitable conflict.  

The estimated role generating process for Consolidators' ego-networks shows a tendency towards high-degree nodes and triadic closure.  This is consistent with \textit{a priori} expectation as Consolidators may use alliances to create communities for the purpose of easing political or economic tie consolidation among all members.  The pooled TERGMs also reveal a tendency towards regime type heterophily within these ego-networks, which makes sense as alliances are often used to build trust \citep{kydd2001trust} or produce converging interests through socialization \citep{bearce2007intergovernmental}.  Existing theory may also expect that asymmetric alliances provide a means of producing concessions between states of disparate capabilities through allowing both states to extract distinct gains through side payments \citep{morrow1991alliances, fordham2010trade}.  We find support for this function, as it appears that Consolidators' ego-networks have a tendency towards alliances formed between states of disparate national capabilities.  

Finally, a pooled TERGM is fit on all ego-networks associated with the Reformers role.  Results from this model are largely consistent with the preceding discussion.  These ego-networks show a tendency towards high-degree nodes and triadic closure.  This would indicate support for the proposition that Reformers seek to include many states in a tightly knit institution, which may assist in socializing member states \citep{pevehouse2005democracy}.  In addition, it is expected that the alliances constituting Reformers' ego-networks are more likely to be asymmetric with respect to capabilities.  This is because stronger states may be more able to effectively exercise influence over weaker states when pushing for domestic reforms \citep{morrow1991alliances}.  Finally, it would be expected that the alliances formed by Reformers are between states of different regimes.  Democracies may more effectively socialize autocratic peers as they'd be seen as more legitimate \citep{pevehouse2005democracy}.  In this case, the positive effect for regime homophily is not consistent with theoretical expectation.  One potential explanation is that reforms have already been implemented, and the model is observing the post-reform tendency for these institutions to persist even after reforms have been made.  
	
	
	

\begin{table}
\begin{center}
\begin{tabular}{l c c c c }
\hline
 & Balancers & Aggregators & Consolidators & Reformers \\
\hline
Edges                          & $\mathbf{-5.95}^{*}$ & $\mathbf{-5.93}^{*}$ & $\mathbf{-1.91}^{*}$ & $\mathbf{-6.22}^{*}$ \\
                               & $[-6.31;\ -5.63]$    & $[-6.69;\ -4.87]$    & $[-2.18;\ -1.64]$    & $[-6.82;\ -5.64]$    \\
Triangles                      & $\mathbf{0.97}^{*}$  & $\mathbf{0.60}^{*}$  & $\mathbf{0.40}^{*}$  & $\mathbf{0.53}^{*}$  \\
                               & $[0.88;\ 1.08]$      & $[0.45;\ 0.76]$      & $[0.35;\ 0.50]$      & $[0.42;\ 0.61]$      \\
GW Degree (0.1)                & $\mathbf{7.86}^{*}$  & $\mathbf{3.71}^{*}$  & $\mathbf{3.31}^{*}$  & $\mathbf{19.18}^{*}$ \\
                               & $[7.19;\ 8.89]$      & $[0.34;\ 6.45]$      & $[2.86;\ 3.96]$      & $[14.65;\ 24.27]$    \\
CINC Difference                & $\mathbf{13.24}^{*}$ & $2.42$               & $\mathbf{8.52}^{*}$  & $\mathbf{15.93}^{*}$ \\
                               & $[12.20;\ 14.30]$    & $[-41.46;\ 13.85]$   & $[6.81;\ 10.25]$     & $[13.67;\ 18.44]$    \\
Revisionism Difference         & $\mathbf{0.13}^{*}$  & $\mathbf{-0.37}^{*}$ & $\mathbf{-0.23}^{*}$ & $-0.02$              \\
                               & $[0.03;\ 0.22]$      & $[-0.68;\ -0.08]$    & $[-0.34;\ -0.13]$    & $[-0.18;\ 0.14]$     \\
Regime Homophily               & $\mathbf{0.72}^{*}$  & $\mathbf{0.54}^{*}$  & $\mathbf{-0.60}^{*}$ & $\mathbf{0.71}^{*}$  \\
                               & $[0.62;\ 0.81]$      & $[0.43;\ 0.65]$      & $[-0.76;\ -0.44]$    & $[0.53;\ 0.89]$      \\
Defensive Commitments          & $\mathbf{1.34}^{*}$  & $\mathbf{1.31}^{*}$  & $\mathbf{2.11}^{*}$  & $\mathbf{2.79}^{*}$  \\
                               & $[1.14;\ 1.55]$      & $[0.57;\ 1.94]$      & $[1.82;\ 2.39]$      & $[2.43;\ 3.23]$      \\
Offensive Commitments          & $\mathbf{1.12}^{*}$  & $\mathbf{2.59}^{*}$  &                      &                      \\
                               & $[0.90;\ 1.36]$      & $[1.06;\ 19.87]$     &                      &                      \\
Secret Provisions              & $\mathbf{0.76}^{*}$  &                      &                      &                      \\
                               & $[0.53;\ 0.95]$      &                      &                      &                      \\
Degree of Institutionalization & $0.30$               & $-3.16$              & $0.01$               & $-0.32$              \\
                               & $[-0.10;\ 0.67]$     & $[-4.21;\ 4.96]$     & $[-0.29;\ 0.22]$     & $[-1.23;\ 0.43]$     \\
Alliance Years                 & $\mathbf{0.06}^{*}$  & $\mathbf{1.24}^{*}$  & $\mathbf{0.02}^{*}$  & $\mathbf{0.06}^{*}$  \\
                               & $[0.04;\ 0.07]$      & $[0.77;\ 1.99]$      & $[0.01;\ 0.03]$      & $[0.03;\ 0.10]$      \\
\hline
Num. obs.                      & 463658               & 612609               & 127956               & 269440               \\
\hline
\multicolumn{5}{l}{\scriptsize{$^*$ 0 outside the 95\% bootstrapped confidence interval}}
\end{tabular}
\caption{\textbf{Pooled TERGM Results for Role Generative Structure, All Periods.}  95\% confidence intervals estimated through bootstrapped pseudolikelihood.   500 replications.}
\label{roleTERGMs}
\end{center}
\end{table}

\section{Concluding Thoughts}
In this manuscript a novel role-based approach to considering alliances is introduced.  This framework breaks with the restrictive and conventional assumption that alliances are homogenous institutions formed in response to external threats.  It also provides a logic necessary to empirically evaluate the rich, nuanced based view of alliances championed by diplomatic historians and security studies scholars. My framework recognizes that states have a variety of objectives that they consider when forming interstate military institutions.  Results indicate that there are many roles made available to states in the alliance network.  These roles, which are informed by contextual factors, influence how and why states form alliances.  This framework and its roles, inferred by the flexible and novel ego-TERGM, offers analysts a new take on perhaps the most foundational concept in IR.  The roles uncovered across distinct historical periods indicate that there is great heterogeneity in the generative model of alliances, and that the motives driving states to form alliances vary across states, space, and time.   

The roles detected by this routine confirm that our initial understanding of military alliances has long been flawed.  This may indicate that many of our theories and empirical models of alliances have been misspecified because they have failed to account for why the alliances were formed in the first place.  Consider any study on alliances, whether it be on the effect of alliances on trade, conflict, or democratization.  By assuming that alliances are formed for the same purpose, empirical models have problematically pooled alliances formed by Balancers or Aggregators with those formed by Reformers or Consolidators and assumed that their effects are constant.  This misspecification plagues both theoretical and empirical work by failing to properly specify the necessary scope conditions about the types of alliances that may be relevant.  

This new framework offers a fresh take on many old and important questions.  Instead of just thinking about the dynamics of different commitments, IR scholars may now empirically consider the dynamics associated with different alliance roles.  In other words, this new framework allows us to ask a variety of new questions: \textit{Are alliances formed by Balancers to restrain peer competitors successful? When Aggregators form alliances to prepare for conflict, do they deter it, or produce it? Are Consolidators successful in using alliances to consolidate economic or political ties? When Reformers use alliances to promote economic or political liberalization, do their allies liberalize? How heterogeneous is the international system?  How does a state's engagement with world politics evolve?}

\clearpage

\clearpage
\bibliographystyle{IO_bibstyle.bst}
\bibliography{AllianceHeterogeneityBib}

\end{document}